\documentclass{article}
\usepackage{graphicx}
\title{\bf Fundamental theories in a phantom universe}
\author{Pedro F. Gonz\'{a}lez-D\'{\i}az\footnote{E-mail:
p.gonzalezdiaz@imaff.cfmac.csic.es}.\\
Colina de los Chopos, Instituto de Matem\'{a}ticas y F\'{\i}sica
Fundamental\\ Consejo Superior de Investigaciones Cient\'{\i}ficas\\
Serrano 121, 28006 Madrid, SPAIN\\ }
\date{November 16, 2004}
\begin{document}
\maketitle \large \setlength{\baselineskip}{0.9cm}

\begin{center}
{\bf Abstract}
\end{center}
Starting with the holographic dark energy model of Li it is shown
that the holographic screen at the future event horizon is sent
toward infinity in the phantom energy case, so allowing for the
existence of unique fundamental theories which are mathematically
consistent in phantom cosmologies.

\vspace{.3cm}


\vspace{.3cm}


\vspace{5.5 cm}

\pagebreak

The existence of the holographic bound in entropy [1] is
manifested also in accelerating cosmology in the form described in
the so-called holographic dark energy models. In fact, recently Li
[2] has proposed a model of holographic dark energy where the
quantum idea that the short distance cut-off is related to the
infrared cut-off [3] is used by assuming that the latter cut-off
is the size of the observer-dependent future event horizon. This
choice allows for a dark energy cosmological scenario and also for
an equation of state of the resulting holographic model which can
be compatible with present observational constraints [4].

If we assume dark energy domination, then the Li model for
holographic dark energy [2] leads to key expressions for the
Friedmann equation and the size of the future event horizon which
are given by
\begin{equation}
H^2 = 8\pi G \rho/3=c^2L^{-2}
\end{equation}
and
\begin{equation}
L=R_h =a\int_t^{\infty}\frac{dt}{a} ,
\end{equation}
where $H=\dot{a}/a$ is the Hubble parameter, with $a\equiv a(t)$
the scale factor, $\rho$ is the dark energy density, $c$ is a
numerical parameter, $L$ is the size of a ultimate region whose
total energy was chosen not to exceed the mass of a black hole
with the same size, and $R_h$ is the size of the future event
horizon. Using Eqs. (1) and (2), with $L=R_h$, Li showed [2] that
the index of the perfect-fluid equation of state $w=p/\rho$, where
$p$ is the pressure, can be written as
\begin{equation}
w= -\frac{1}{3}-\frac{2}{3c} ,
\end{equation}
in which the parameter $c$ can generally take on values $\geq 1$ and
also $< 1$. If $c > 1$ then we have a holographic quintessence
model, if $c=1$ we obtain a cosmological description that
corresponds to the existence of a positive cosmological constant,
and if $c<1$ we obtain a phantom cosmology model [5]. Although Li
argued in favor of the cosmological constant case in Ref. [2], any
value of $c$ is in principle allowed by the model.

In this short letter we shall show that whereas the size of the
future event horizon allowed by the Li model is always finite for
$c\geq 1$, for $c<1$ $R_h$ should become necessarily infinite,
contrary to the claim of Li himself [2]. Our argument is almost
immediate. We start with the general expression for the dark energy
density derived from the conservation law for cosmic energy,
$\rho=\propto a^{-2(1-1/c)}$, which, when inserted into the first
equality in the Friedmann equation (1), leads to a general
expression for the scale factor given by
\begin{equation}
a= \left(a_0^{1-1/c}+ (1-\frac{1}{c})(t-t_0)\right)^{1/(1-1/c)} ,
\end{equation}
with $a_0$ the value of the scale factor at the initial time $t_0$.
Although an initial value of the scale factor can be set equal to 1,
for the sake of generality we shall keep an arbitrary value for it.
Thus, inserting Eq. (4) into Eq. (2) and integrating, we obtain for
the future event horizon the general expression
\begin{equation}
R_h
=-cT^{1/(1-1/c)}\left(\left.T^{-1/[c(1-1/c)]}\right|_{t}^{\infty}\right)
,
\end{equation}
where
\begin{equation}
T=a_0^{1-1/c}+ (1-\frac{1}{c})(t-t_0)\equiv a^{1-1/c}.
\end{equation}

It can first be checked that the holographic expression $H^2=c^2
R_h^{-2}$ is in fact satisfied when we use Eqs. (4) and (5) for any
value of parameter $c$. Next, we can readily see that if $c
>1$ then $R_h=cT$, but when we allow $c < 1$, then inexorably $R_h
=\infty$\footnote{Note that $T$ becomes negative after the big rip
[8] at $t_* =t_0+[a_0^{1/c-1)}(1/c-1)]^{-1}$}. Thus, the
holographic phantom energy model directly constructed from the
general holographic dark energy model of Li by simply imposing the
condition $c<1$ sends the future event horizon towards
infinity\footnote{Since at the big rip there is a real singularity
it could be thought that the whole universe after the big rip
would be inaccessible to current observers, as one should cut the
infinite hypersurface at $t=t_*$, leaving the universe at $t>t_*$
disconnected from such observers. However, sufficiently grown up
Lorentzian wormholes are made possible in the phantom regime which
may connect the regions before and after the big rip to one
another [9], so rendering connected the whole spacetime. Had we
cut off all the space-time after the big rip, then the region with
size $L$ ought to be replaced for a new horizon at time $t=t_*$ in
the future, defined by
\[R_*=a\int_t^{t_*}dt/a=cT=R_h(c<1)<R_h(c\geq 1) ,\]
which, at first sight, would even aggravate the event horizon
puzzle relative to the quintessence and cosmological constant
cases. Thus, if no wormholes were allowed to occur, an
interpretation in terms of cosmological complementarity would
again be required (see later on). Becoming the accessible universe
then infinite though will make any fundamental theories based on a
S-matrix or S-vector formulation to be consistently defined
mathematically.}. This result has an immediate implication in
string and M theories. In fact, it follows from the above
discussion that whereas the puzzle about how a fundamental theory
can be formulated within a finite box in accelerating cosmology
[6] remains operative in holographic quintessence or cosmological
constant models, such a puzzle automatically disappears in
holographic phantom cosmology as all the space regions are in this
case fully accessible to light probes and therefore the S-matrix
or S-vector description required in string or other fundamental
theories could perfectly be mathematically formulated.

We interpret this result as follows. The question of how a
fundamental theory can be constructed within the finite region
enclosed by the future event horizon of de Sitter and quintessence
cosmological models has been regarded to be a consequence of the
so-called cosmological complementarity [7] from which the
wave/particle duality or the superposition principle of quantum
mechanics are thought to be nothing but mere limiting cases.
According to the cosmological duality principle all objects in the
universe have two aspects of their existence, on the one hand they
can behave as members of the whole cosmic collective and, on the
other hand, such objects can also behave as local individuals in
the universe, so that all these objects ought then to be defined
by pairs of numbers, each pair specifying the value of the
dynamical physical variable describing the state of the object. It
appears clear that the algebra of complex numbers may account for
this, whereby the real component is associated with the collective
gravitational aspect and the imaginary component is associated
with the individualized aspect describing the forces that have to
be accommodated within a common gauge theory. In the accelerating
cosmological scenario with $c\geq 1$ there can then be many
S-matrices of fundamental theories which are related to each other
by gauge transformations and physically correspond to the
superposition implied by the principle of cosmological
complementarity. Now, the emergence in the future of a big rip
singularity in phantom cosmologies [8] (see however Ref. [10])
breaks this duality, leaving only a cosmic behavior for all matter
in the universe, at least as one approaches the singularity. It
follows that in phantom cosmologies showing a big rip singularity
in the future and allowing for growing Lorentzian
wormholes\footnote{The main difficulty which has been encountered
for allowing the existence of wormholes is the quantum instability
produced by catastrophic vacuum fluctuations near their throat
[13]. Nevertheless, if as one approaches the big rip singularity
the quantum-mechanically sustained microscopic, local behaviour of
any objects is being erased off, then the wormholes would
stabilize near the singularity and the universe can whereby become
connected throughout its entire evolution} it is possible to have
a unique mathematically well-defined S-matrix or S-vector
description and hence a unique fundamental theory. It is for this
reason that in the field theory description of a phantom fluid,
one has necessarily to introduce a Wick rotation of the phantom
field which converts this field into a gravitational object, and
makes negative the scalar-field kinetic term [11]. Such negative
kinetic terms are though the ultimate responsible for the the
existence of the violent instabilities found for the phantom
fluids [12]

\vspace{.8cm}

\noindent{\bf Acknowledgements} The author thanks Carmen L.
Sig\"{u}enza and A. Rozas Fern\'{a}ndez for useful information
exchange and comments. This work was supported by DGICYT under
Research Project No. BMF2002-03758.

\pagebreak

\noindent\section*{References}

\begin{description}

\item [1] P.F. Gonz\'{a}lez-D\'{\i}az, Phys. Rev. D27 (1983) 3042;
G. 't Hooft, in: {\it Salam-festschrifft}, edited by A. Aly, J.
Ellis, S. Randjbar-Daemi (World Scientific, 1993); L. Susskind, J.
Math. Phys. 36 (1995) 6377 .
\item [2] M. Li, {\it A Model for Holographic Dark Energy},
hep-th/0403127
\item [3] A. Cohen, D. Kaplan and A. Nelson, Phys. Rev. Lett. 82 (1999)
4971 .
\item [4] A.G. Riess {\it et al.},Astrophys J. 607 (2004) 665 .
\item [5] R.R. Caldwell, Phys. Lett. B545 (2002) 23 .
\item [6] T. Banks, {\it Cosmological Breaking of Supersymmetry},
hep-th/0007146; E. Witten, {\it Quantum Gravity in De Sitter Space},
hep-th/0106109; X.-G. He, {\it Accelerating Universe and Event
Horizon}, astro-ph/0105005 .
\item [7] T. Banks and W. Fischler, {\it M Theory Observables for
Cosmological Space-Times}, hep-th/0102077 .
\item [8] R.R. Caldwell, M. Kamionkowski and N.N. Weinberg, Phys.
Rev. Lett. 91 (2003) 071301; S. Nojiri and S.D. Odintsov, Phys.
Lett. B562 (2003) 147; L.P. Chimento and R. Lazkoz, Phys. Rev.
Lett. 91 (2003) 211301; S. Nesseris and L. Perivolaropoulos, {\it
The Fate of Bound Systems in Phantom and Quintessence
Cosmologies}, astro-ph/0410309 .
\item [9] P.F. Gonz\'{a}lez-D\'{\i}az, Phys. Rev. D68 (2003) 084016;
Phys. Rev. Lett. 93 (2004) 071301 .
\item [10] P.F. Gonz\'{a}lez-D\'{\i}az, Phys. Rev. D68 (2003) 021303;
M. Bouhmadi-L\'{o}pez and J.A. Jim\'{e}nez Madrid, {\it Escaping the
Big Rip}, astro-ph/0404540 .
\item [11] P.F. Gonz\'{a}lez-D\'{\i}az, Phys. Rev. D69 (2004) 063522
.
\item [12] S. M. Carroll, M. Hoffman and M. Trodden, Phys. Rev. D68
(2003) 023509 .
\item [13] S.W. Hawking, Phys. Rev. D46 (1992) 603 .

\end{description}

\end{document}